\documentclass[journal]{IEEEtran}
\usepackage{amsmath,graphicx,url,times}
\usepackage{color}
\usepackage{lipsum}
\usepackage[numbers,sort&compress]{natbib}

\pdfoutput=1

\ifCLASSOPTIONcompsoc
\usepackage[caption=false,font=normalsize,labelfon
t=sf,textfont=sf]{subfig}
\else
\usepackage[caption=false,font=footnotesize]{subfi
g}
\fi

\usepackage{enumitem}
\usepackage{booktabs,tabularx}
\newcolumntype{C}{>{\centering\arraybackslash}X} 
\usepackage{amsmath,amsfonts,bm,amsthm}
\usepackage{mathtools}
\usepackage{color}
\usepackage{physics}
\usepackage{nth}
\usepackage{todonotes}

\usepackage{xparse}
\DeclarePairedDelimiterX{\Set}[1]{\{}{\}}{\setargs{#1}}
\NewDocumentCommand{\setargs}{>{\SplitArgument{1}{;}}m}
{\setargsaux#1}
\NewDocumentCommand{\setargsaux}{mm}
{\IfNoValueTF{#2}{#1} {#1\,\delimsize|\,\mathopen{}#2}}
\DeclarePairedDelimiter\paren{\lparen}{\rparen}

\DeclarePairedDelimiter\bracket{[}{]}

\DeclarePairedDelimiter\euclidean{\lVert}{\rVert_2}

\renewcommand{\vec}[1]{\bm{#1}}
\newcommand{\mat}[1]{\bm{#1}}
\newcommand{\tran}{^{\mathstrut\scriptscriptstyle\top}} 
\newcommand{\der}{'}
\newcommand{\herm}{^{\mathstrut\scriptscriptstyle H}} 
\newcommand{\conj}{*}
\newcommand{\diag}[1]{\mathrm{diag}\paren*{#1}}

\newcommand{\adj}[0]{\mathrm{adj}}

\newcommand{\detp}[1]{\det\paren*{#1}}
\newcommand{\cond}[0]{\kappa}
\newcommand{\eye}[0]{\mat{I}}
\newcommand{\ones}[0]{\vec{1}}

\newcommand{\BigO}[0]{\mathcal{O}}

\newcommand{\stateSpace}[1]{\mathfrak{#1}}

\newcommand{\A}[0]{{\mat{\stateSpace{A}}}}

\newcommand{\N}[0]{\stateSpace{{N}}}

\newcommand{\tf}[1]{h_\textrm{#1}}
\newcommand{\tfn}[1]{h_{#1}}
\newcommand{\Tf}[1]{H_\textrm{#1}}

\newcommand{\set}[1]{\mathbb{#1}}

\newcommand{\gcp}[0]{p}
\newcommand{\gcq}[0]{q}
\newcommand{\pole}[0]{\lambda}
\newcommand{\residue}[0]{\rho}
\newcommand{\unresidue}[0]{\residue^\mathrm{u}}

\newcommand{\sv}[0]{\sigma}

\newcommand{\delay}[0]{m}
\newcommand{\Delay}[0]{\vec{\delay}}
\newcommand{\atten}[0]{\Gamma}

\newcommand{\Fbm}[0]{\mat{A}}

\newcommand{\directgain}[0]{d}
\newcommand{\Ingain}[0]{\vec{b}}
\newcommand{\Outgain}[0]{\vec{c}}

\newcommand{\spectralRadius}[0]{\rho}

\newcommand{\matSize}[0]{N}
\newcommand{\stdDelay}[0]{\mat{D}_{\Delay}(z)}

\newcommand{\stdDelayArg}[1]{\mat{D}_{\Delay}(#1)}

\newtheorem{theorem}{Theorem}

 \usepackage{mathtools}

 \newcommand{\RT}[0]{T_{60}} 
 \newcommand{\FS}[0]{f_{s}} 
 
\graphicspath{{./}}

\usepackage{algorithm}
\usepackage[noend]{algpseudocode}

\algrenewcommand\algorithmicrequire{\textbf{Input:}}
\algrenewcommand\algorithmicensure{\textbf{Output:}}

\newcommand{\matPoly}[0]{P}
\newcommand{\MatPoly}[0]{\mat{P}}
\newcommand{\tol}[0]{\tau}
\newcommand{\newton}[1]{\mathcal{N}\paren*{#1}} 
\newcommand{\newtonS}[0]{\mathcal{N}} 
\newcommand{\EAIstep}[2]{\Delta_{#1}\iter{#2}} 
\newcommand{\approxEAIstep}[2]{\widetilde{\Delta}_{#1}\iter{#2}} 
\newcommand{\deflation}[2]{\mathcal{D}_{#1}\paren*{#2}} 
\newcommand{\deflationS}[1]{\mathcal{D}_{#1}} 
\newcommand{\approxDeflation}[2]{\widetilde{\mathcal{D}_{#1}}\paren*{#2}} 
\newcommand{\approxDeflationS}[1]{\widetilde{\mathcal{D}_{#1}}} 
\newcommand{\deflationError}[0]{\err_\mathcal{D}}
\newcommand{\iter}[1]{^{(#1)}} 
\newcommand{\outsignal}[0]{y}
\newcommand{\insignal}[0]{x}
\newcommand{\state}[0]{s}
\newcommand{\StateVec}[0]{\vec{\state}}
\newcommand{\err}[0]{\epsilon}

\newcommand{\shiftMatrix}[0]{\eye_\textrm{S}}
\newcommand{\numNear}[0]{\N_\textrm{near}}
\newcommand{\clusterNumber}[0]{\mathcal{C}}
\newcommand{\poleIndex}[0]{i}
\newcommand{\iterIndex}[0]{j}
\newcommand{\polyDegree}[0]{K}
\newcommand{\polyIndex}[0]{k}
\newcommand{\hist}[0]{\mathcal{C}_{\textrm{HIST}}}
\newcommand{\histIndex}[0]{\kappa}
\newcommand{\clusterProbe}[0]{\mathcal{L}}

\title{Modal Decomposition of Feedback Delay Networks}

\author{Sebastian J. Schlecht and 
        Emanu\"el A. P. Habets,~\IEEEmembership{Senior Member,~IEEE}
        }
        
\begin{document}

\maketitle

\begin{sloppy}

\begin{abstract}
Feedback delay networks (FDNs) belong to a general class of recursive filters which are widely used in sound synthesis and physical modeling applications. We present a numerical technique to compute the modal decomposition of the FDN transfer function. The proposed pole finding algorithm is based on the Ehrlich-Aberth iteration for matrix polynomials and has improved computational performance of up to three orders of magnitude compared to a scalar polynomial root finder. We demonstrate how explicit knowledge of the FDN's modal behavior facilitates analysis and improvements for artificial reverberation. The statistical distribution of mode frequency and residue magnitudes demonstrate that relatively few modes contribute a large portion of impulse response energy.    
\end{abstract}


\section{Introduction}
\label{sec:intro}


\IEEEPARstart{A}{} feedback delay network (FDN) consists of a set of delay lines with lengths $\Delay$ which are interconnected via a feedback matrix $\Fbm$ (see Fig.~\ref{fig:ModalDecompositionConcept}). FDNs arise in many physical modeling applications where geometrically distributed components are approximated by time delays, e.g., strings \cite{Karplus:1983bq}, plates and membranes \cite{SmithIII:1992bn}, springs \cite{Parker:2011fn}, and air volume \cite{Savioja:2015ft,Valimaki:2012jv}. The interest in FDNs is fueled by the highly efficient implementation of delays in the time-domain, e.g., with circular buffers resulting in a constant time complexity $\BigO(1)$ independent of its length. Therefore, the computational complexity of the FDN scales with the number of delay lines and not the system order. FDNs are a popular choice for artificial reverberation applications particularly because of the favorable relation between FDN size and system order \cite{Gerzon:1971tu,Jot:1991tq,Schlecht:2017il,Schlecht:2017jt}.

In this work, we present a modal decomposition technique for FDNs. The modal decomposition of a system is an equivalent representation as the sum of complex one-pole resonators, so-called modes. The time-domain signal of such a resonator with pole $\pole_\poleIndex$ and residue $\residue_\poleIndex$ is
\begin{equation}
	\tfn{\poleIndex}(n) = \abs{\residue_\poleIndex} \abs{\pole_\poleIndex}^n \, e^{ \imath (n \angle \pole_\poleIndex + \angle \residue_\poleIndex) },
	\label{eq:modeExp}
\end{equation}
where $\angle$ indicates the argument of a complex number in radiant, $\abs{\cdot}$ is the magnitude, $\imath = \sqrt{-1}$ and $n$ indicates the discrete time index. Each individual resonating mode is governed by four parameters: mode frequency $\angle \pole_\poleIndex$, decay rate $\abs{\pole_\poleIndex}$, initial phase $\angle \residue_\poleIndex$ and initial amplitude $\abs{\residue_\poleIndex}$ (see Fig~\ref{fig:ModalDecompositionConcept}). With modal decomposition, we aim to uncover the specific parameters of each mode. The time-domain impulse response of the FDN
\begin{equation}
	\tf{}(n) = \sum_{i=1}^\N \tfn{\poleIndex}(n)
	\label{eq:sumModes}
\end{equation}
is the sum of the complex modes $\tfn{\poleIndex}(n)$, where $\N$ is the system order. 
In sound synthesis applications for instance, the human auditory system can recognize the spectral quality composed of the individual modes and this representation is therefore termed additive or modal synthesis \cite{Adrien:1991ub}. Modal analysis of recursive systems is applied in various system modeling applications, ranging from acoustics and digital filter design to mechanical modeling \cite{He:2001cw}. A particularly challenging application for modal decomposition is room acoustics, where even medium room sizes exhibits millions of modes \cite{Kuttruff:2009vl}. Only for simple room geometries, an analytic expression for the system poles and residues can be stated \cite{Naka:2005bd}. System poles may also be recovered from the impulse response by various techniques such as an autoregressive moving-average \cite{Karjalainen:2002wk}, Bayesian inference \cite{Beaton:2017jk} and all-pole modeling \cite{Haneda:1994fi}. Whereas these techniques may be able to successfully compute partial solutions or compute the solution for specific configurations, the computation of the entire set of modes is in general challenging. In the following, we give the precise problem statement of this work.

\begin{figure}[!t]
  \includegraphics[]{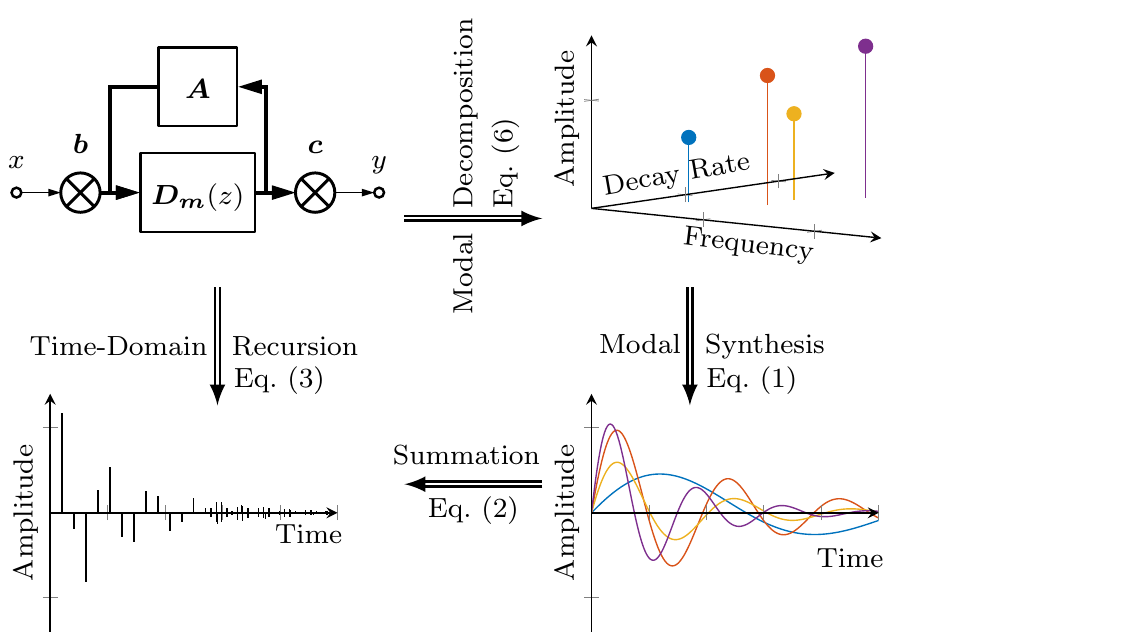}
  \caption{Conceptual overview of modal decomposition and synthesis of a feedback delay network (FDN). \emph{Top left}: FDN block diagram with a set of delay lines $\stdDelay$, connected via a feedback matrix $\Fbm$, and input and output gains $\Ingain$ and $\Outgain$ for input and output signals $x$ and $y$, respectively. Thick lines indicate multiple signals. \emph{Top right}: FDN modes with four parameters each: frequency, decay rate, initial amplitude and phase (not depicted). \emph{Bottom right}: Time-domain impulse responses of the resonators corresponding to the FDN modes. \emph{Bottom left}: Time-domain impulse response of the FDN.}
  \label{fig:ModalDecompositionConcept}
\end{figure}

\subsection{Problem Statement}
For a single input and single output, the time-domain recursion of an FDN with $\matSize$ delay lines is given by
\begin{equation}
\begin{aligned}
	\outsignal(n) = \Outgain\tran \StateVec(n) + \directgain \insignal(n)	\\
	\StateVec(n + \Delay) = \Fbm \StateVec(n) + \Ingain \insignal(n),
\end{aligned}
	\label{eq:timeDomainFDN}
\end{equation}
where $n$ is the time index, $\cdot\tran$ denotes the transpose operation and $\Fbm \in \set{C}^{\matSize \times \matSize}$, $\Ingain, \Outgain, \StateVec(n) \in \set{C}^{\matSize \times 1}$, $\insignal(n), \outsignal(n), \directgain \in \set{C}$ \cite{Rocchesso:1997fv}. The state vector is defined as $\StateVec(n + \Delay) = \bracket*{ \state_1(n + \delay_1), \dots, \state_N(n + \delay_N)}$. We write $\matSize$-FDN to denote an FDN of size $\matSize$. The transfer function of an FDN is 
\begin{equation}
	\Tf{}(z) = \Outgain\tran \bracket*{\stdDelay^{-1} -  \Fbm}^{-1} \Ingain + \directgain,
	\label{eq:transferFunction}
\end{equation}
where $\stdDelay = \diag{ z^{-\delay_1}, z^{-\delay_2}, \dots, z^{-\delay_\matSize}}$. The system order is given by $\N = \sum_{i=1}^\matSize \delay_i$ \cite{Rocchesso:1997fv}. For commonly used delays $\Delay$, the system order is much larger than the FDN size, i.e.,
\begin{equation}
	\N \gg \matSize.
	\label{eq:longDelays}
\end{equation}
The modal decomposition of the FDN, i.e., the partial fraction decomposition (PFD) of the transfer function \eqref{eq:transferFunction} is
\begin{equation}
	H(z) = \directgain + \sum_{\poleIndex=1}^\N \frac{\residue_\poleIndex}{1 - \pole_\poleIndex \,z^{-1}},
	\label{eq:modalDecomposition}
\end{equation}
where $\residue_\poleIndex$ is the residue of the pole $\pole_\poleIndex$. The time-domain representation of the sum in \eqref{eq:modalDecomposition} is given in \eqref{eq:sumModes} as the sum of complex resonators. The objective of this work is to present an efficient numerical method to compute the modal decomposition \eqref{eq:modalDecomposition} from the transfer function \eqref{eq:transferFunction}.

\subsection{Direct Approach}
\label{sec:naive}
We first review two standard methods for the modal decomposition \cite{Rocchesso:1997fv,Schlecht:2015hi}. Let $\Fbm$ be any invertible matrix, then 
\begin{equation}
	 \adj(\Fbm) = \detp{\Fbm} \Fbm^{-1},
	\label{eq:adjugate}
\end{equation}
where $\adj(\Fbm)$ is the adjugate of the matrix $\Fbm$ \cite{Golub:1996wp}. In the following, we denote
\begin{equation}
	\MatPoly(z) = \stdDelay^{-1} -  \Fbm.
	\label{eq:loopTransfer}
\end{equation}
With \eqref{eq:adjugate} and \eqref{eq:loopTransfer}, the transfer function \eqref{eq:transferFunction} can be expressed as a rational polynomial 
\begin{equation}
	\Tf{}(z) = \frac{\gcq_{\Delay, \Fbm, \Ingain, \Outgain, \directgain}(z)}{\gcp_{\Delay, \Fbm}(z)}, 
\end{equation}
where
\begin{equation}		
	\gcp_{\Delay, \Fbm}(z) = \detp{ \MatPoly(z) } 
	\label{eq:gcp}
\end{equation}	
and
\begin{equation}
\begin{aligned}
	\gcq_{\Delay, \Fbm, \Ingain, \Outgain, \directgain}(z) =
	\directgain \detp{ \MatPoly(z) } + \Outgain\tran \, \adj(\MatPoly(z)) \, \Ingain
	\label{eq:gcq}.
\end{aligned}
\end{equation}
For brevity, we occasionally omit the parameters and write $\gcq(z)$ and $\gcp(z)$. The FDN system poles $\pole_\poleIndex$, where $1 \leq i \leq \N$, are the roots of the generalized characteristic polynomial (GCP) $\gcp_{\Delay, \Fbm}(z)$ in \eqref{eq:gcp} such that they are fully characterized by the delay matrix $\stdDelay$ and the feedback matrix $\Fbm$.

For a moment, let us assume that all delays are single time steps, i.e., $\Delay = \ones$. The time-domain recursion in \eqref{eq:timeDomainFDN} reduces to the standard state-space description of a linear time-invariant (LTI) filter. The system poles $\pole_\poleIndex$ are the eigenvalues of the feedback matrix $\Fbm$ such that the modal decomposition \eqref{eq:modalDecomposition} is easily computed with standard methods. However, for longer delays $\Delay$ such that \eqref{eq:longDelays} holds, the modal decomposition becomes more involved.

The GCP $\gcp_{\Delay, \Fbm}$ can be expressed in a linearized fashion 
\begin{equation}
	\gcp_{\Delay, \Fbm}(z) = \detp{ z \eye_{\N} - \A },
	\label{eq:linearization}
\end{equation}
where $\mat{I}_{\N}$ is the identity matrix of size $\N$ and $\A \in \set{C}^{\N \times \N}$ such that the system poles are the eigenvalues of $\A$ \cite{Rocchesso:1997fv}. Unfortunately, for large delays $\Delay$ this eigenvalue problem becomes quickly numerically intractable. Alternatively, the GCP can be expressed as a scalar polynomial 
\begin{equation}
	\gcp_{\Delay, \Fbm}(z) = \sum_{i=0}^\N c_i z^{i},
	\label{eq:gcp_pm}
\end{equation}
where the coefficients $c_i$ are derived from the principal minors of $\Fbm$ \cite{Schlecht:2015hi}. The system poles are the roots of the scalar polynomial. Again, the polynomial degree increases with longer delays $\Delay$ and finding the roots of the polynomial becomes numerically intractable \cite{McNamee:2007vi}. 


In the remainder of this paper, we present a numerically stable and computationally efficient method to compute the modal decomposition for large system order $\N$ and modest-sized $N$. In Section~\ref{sec:numericalModalDecomposition}, we derive the fundamental algorithm based on a polynomial matrix formulation. In Section~\ref{sec:modalEvaluation}, we evaluate the performance of the proposed algorithm. In Section~\ref{sec:analysisFDN}, we apply modal decomposition to analyze the effects of attenuation filters and to study the statistical distributions of mode frequencies and residue magnitudes.

\section{Numerical Modal Decomposition}
\label{sec:numericalModalDecomposition}
In the following, we present a root finding algorithm for the GCP $\gcp(z)$ and subsequently recover the residues $\residue_\poleIndex$. We conclude this section with a generalization to additional filtering in the delay lines and feedback matrix.


\subsection{Polynomial Matrix Formulation}
It is a common heuristic in numerical computation that the inherent problem structure shall be preserved as much as possible throughout all computation steps to improve numerical performance. In contrast to Section~\ref{sec:naive}, we compute the system poles without expanding the problem. In fact, \eqref{eq:gcp} is a polynomial eigenvalue problem of degree $\polyDegree = \max \Delay$, i.e.,
\begin{equation}
	\MatPoly(z) = \sum_{\polyIndex=0}^\polyDegree \MatPoly_\polyIndex \, z^{\polyIndex},
	\label{eq:polyEigenvalue}
\end{equation}
where $\MatPoly_\polyIndex \in \set{C}^{N \times N}$ for $0 \leq \polyIndex \leq \polyDegree$. For a proper matrix polynomial $\MatPoly(z)$, i.e., $\detp{ \MatPoly_\polyDegree }\neq 0$, the number of roots is $\polyDegree \matSize$ \cite{Bini:2013fo}. For FDNs, however, $\MatPoly_\polyDegree$ is singular such that the actual number of roots is lower, respectively, many roots are infinite. In fact, if $\detp{\Fbm} \neq 0$, the number of finite roots is $\N$ which is also the degree of the scalar polynomial in \eqref{eq:gcp} \cite{Schlecht:2015hi}.   
 
In the following, we use the derivative of the polynomial $\gcp(z) = \detp{\MatPoly(z)}$. According to Jacobi's formula \cite{Higham:2012er}, we have
\begin{equation}
\begin{aligned}
	\gcp\der(z) = \dv{}{z} \gcp(z) =& \detp{\MatPoly(z)} \, \trace \paren*{ \MatPoly(z)^{-1} \,\MatPoly'(z) } \\
	=&	\trace \paren*{ \adj(\MatPoly(z)) \,\MatPoly'(z) },
	\label{eq:detDerivative}
\end{aligned}
\end{equation} 
where $\MatPoly'(z) = \dv{\MatPoly(z)}{z}$ and $\trace(\mat{X})$ denotes the trace of matrix $\mat{X}$. Stewart \cite{Stewart:1998gg} showed that the adjugate of $\Fbm$ can be well-conditioned even when $\Fbm$ is ill-conditioned, and he shows how $\adj(\Fbm)$ can be computed in a numerically stable way from a rank revealing decomposition of $\Fbm$ \cite{Higham:2012er}. With the definitions of the polynomial eigenvalue problem introduced, we present the proposed root finding algorithm.

\subsection{Ehrlich-Aberth Method}
The polynomial eigenvalue problem can be solved with the Ehrlich-Aberth Iteration (EAI) method, i.e., a combination of Newton method and a deflation term which prevents that two eigenvalues converge to the same solution \cite{Bini:2013fo}. Let $\vec{\pole}\iter{0} \in \set{C}^{\N}$ be a vector of initial estimates for the $\N$ roots of the polynomial $p(z)$ and $\vec{\pole}\iter{\iterIndex} = \bracket{\pole_1\iter{\iterIndex}, \pole_2\iter{\iterIndex}, \dots, \pole_\N\iter{\iterIndex} }$ be the $\iterIndex$-th EAI iteration. The EAI provides the sequence of estimates
 \begin{equation}
	\pole_\poleIndex\iter{\iterIndex+1} = \pole_\poleIndex\iter{\iterIndex} - \EAIstep{\poleIndex}{\iterIndex}
	\label{eq:EAI}
\end{equation}
with the EAI step being
\begin{equation}
	\EAIstep{\poleIndex}{\iterIndex} = \frac{\newton{\pole_\poleIndex\iter{\iterIndex}}}{1 - \newton{\pole_\poleIndex\iter{\iterIndex}} \deflation{\poleIndex}{\vec{\pole}\iter{\iterIndex}} }.
	\label{eq:EAIstep}
\end{equation}
Using the identity in \eqref{eq:detDerivative}, the Newton correction term is
\begin{equation}
	\newton{z} = \frac{\gcp(z)}{\gcp'(z)} = \frac{1}{ \trace \paren*{ \MatPoly(z)^{-1} \, \MatPoly'(z) } }
	\label{eq:newtonCorrection}
\end{equation} 
and the deflation term is
\begin{equation}
	\deflation{\poleIndex}{\vec{\pole}\iter{\iterIndex}} = \sum_{l=1, l\neq \poleIndex}^\N \frac{1}{\pole_\poleIndex\iter{\iterIndex} - \pole_l\iter{\iterIndex}}.
	\label{eq:Deflation}
\end{equation}
The deflation term may be interpreted as a penality term if two eigenvalues approach each other too closely and guarantees that the all eigenvalues reached are unique. Using \eqref{eq:newtonCorrection} we can expand \eqref{eq:EAIstep} to
\begin{equation}
	\EAIstep{\poleIndex}{\iterIndex} = \frac{1}{\trace \paren*{ \MatPoly \paren*{\pole_\poleIndex\iter{\iterIndex}}^{-1} \, \MatPoly'\paren*{\pole_\poleIndex\iter{\iterIndex}} } - \deflation{\poleIndex}{\vec{\pole}\iter{\iterIndex}} }.
	\label{eq:EAIsimple}
\end{equation}
The method, given here in the Jacobi version, is known to converge cubically for simple roots and linearly for multiple roots \cite{Bini:2013fo}. The Gauss-Seidel version of EAI \cite{Bini:2013fo}, which updates the estimates as soon as they become available, may converge even slightly faster.

\subsection{Stopping Criteria}
The system poles $\pole_\poleIndex$ are the roots of the polynomial $\gcp(z)$ in \eqref{eq:gcp}, i.e., $\detp{ \MatPoly(\pole_\poleIndex) } = 0$. In other words, $\MatPoly(\pole_\poleIndex)$ is a singular matrix for all system poles. Thus, the natural stopping criteria is the reciprocal of the condition number $\cond(\MatPoly(z))$ being less than a prescribed tolerance $\tol_1$. This stopping condition is also computationally favorable as the condition number can be estimated highly efficiently \cite{Higham:2012er}. However, for multiple eigenvalues this stopping condition may result in a premature halt \cite{Bini:2013fo}.

An alternative stopping condition says that the computed correction is too tiny and would not change the significant digits of the current estimate 
\begin{equation}
	\abs{ \EAIstep{\poleIndex}{\iterIndex} } \leq \tol_2 \abs{\pole_\poleIndex\iter{\iterIndex}}, 
\end{equation}
where $\tol_2$ is a small positive tolerance threshold \cite{Bini:2013fo}. In practice, good global convergence properties are observed; a theoretical analysis of global convergence, though, is still missing and constitutes an open problem. There is empirical evidence that the number of Newton iterations heavily depends on the choice of the initial estimates \cite{Bini:2013fo}.

\subsection{Initialization}
Aberth \cite{Aberth:1973ku} proposed to choose initial estimates placed along a circle centered at the origin of sufficiently large radius so that it contains all the roots. In case the magnitude of the roots vary largely, multiple circles with suitable radii may be chosen instead \cite{Bini:1996fk}. With Rouch\'e's theorem, we can derive upper and lower bounds on the pole magnitudes for the FDN depending on the singular values $\sv(\Fbm)$ of the feedback matrix (see Appendix~\ref{sec:magnitudeBounds}) 
\begin{equation}
	\sqrt[\min \Delay]{\min \sv(\Fbm)} \leq \abs{\pole_\poleIndex} \leq \sqrt[\max \Delay]{\max \sv(\Fbm)}. 
	\label{eq:magnitudeBounds}
\end{equation}
Equation~\eqref{eq:magnitudeBounds} is a generalization on the relation of eigenvalues and singular values as given in the Weyl-Horn Theorem \cite{Horn:1954fr} in the case of unit delays $\Delay = \ones$. 
The bound is tight for a diagonal feedback matrix where the minimum and maximum delays coincide with the minimum and maximum diagonal element, respectively. However, the bound may be arbitrarily loose. For instance, the maximum singular value of a triangular matrix $\max \sv(\Fbm)$ may be arbitrarily large while all system poles lie on the unit circle \cite{Schlecht:2017jt}. For large delays $\Delay$ however, \eqref{eq:magnitudeBounds} shows that the pole magnitudes tend to be close to the unit circle. 

We can further derive from \eqref{eq:magnitudeBounds} that if $\max \sv(\Fbm) \leq 1$ then all poles lie on the closed unit disk which is equivalent to the FDN being marginally stable \cite{Proakis:2007fs}. In particular, if all singular values are 1, which is equivalent to $\Fbm$ being unitary, i.e., $\Fbm\herm \Fbm = \eye$, all system poles lie on the unit circle regardless of the delays $\Delay$. Such an FDN is called lossless, and represents an important special case \cite{Schlecht:2017jt}.

In this work, we are interested in lossless and stable FDNs for their practical relevance. In combination with \eqref{eq:longDelays}, such FDNs have all poles in the unit disk, but close to the unit circle. Thus, we place the initial estimates $\vec{\pole}\iter{0}$ uniformly on the unit circle. More precisely, we chose the roots of unity
\begin{equation}
	\vec{\pole}\iter{0} = \exp \paren*{ \imath 2\pi \bracket*{\frac{0}{\N}, \frac{1}{\N}, \dots, \frac{\N-1}{\N} } }.
	\label{eq:initialGuess}
\end{equation}
 It is worthwhile to note that $\vec{\pole}\iter{0}$ is the solution of a particular FDN with a circular shift matrix
\begin{equation}
	\Fbm = \shiftMatrix = \begin{bmatrix}
		0 & 1 & 0 & \cdots & \cdots & 0 \\
		0 & 0 & 1 & \cdots & \cdots & 0 \\
		\vdots & \vdots & \vdots & \ddots & 1 & 0 \\
		0 & 0 & 0 & \cdots & 0 & 1 \\
		1 & 0 & 0 & \cdots & 0 & 0 \\
	\end{bmatrix}
\end{equation}
such that the GCP is
\begin{equation}
	\gcp_{\Delay, \shiftMatrix}(z) = z^\N - 1. 
\end{equation}
The shift matrix thus combines the FDN delays into a single long delay line.


\subsection{Approximate Deflation}
For a high system order $\N$, the computational complexity of the deflation term \eqref{eq:Deflation} may become excessive. We propose an \emph{approximate deflation (AD)} according to a maximum error tolerance $\tol_3$ for the resulting EAI step $\approxEAIstep{\poleIndex}{\iterIndex}$ in \eqref{eq:EAIsimple}, i.e.,
\begin{equation}
	\abs{ \EAIstep{\poleIndex}{\iterIndex}  -
	\approxEAIstep{\poleIndex}{\iterIndex} } 
	\leq \tol_3.
	\label{eq:EAIerror}
\end{equation}
The magnitude of the deflation term summands decreases with the pole distance $\pole_\poleIndex\iter{\iterIndex} - \pole_l\iter{\iterIndex}$. The idea is then to divide the poles into a near and far pole sets, $\vec{\pole}_{\textrm{near}}\iter{\iterIndex}$ and $\vec{\pole}_{\textrm{far}}\iter{\iterIndex}$, respectively, and approximate the deflation of the less significant far poles $\vec{\pole}_{\textrm{far}}\iter{\iterIndex}$ by a default term, e.g., $\vec{\pole}_{\textrm{far}}\iter{0}$. For symmetry, the number of near poles $\numNear$ is assumed to be an even number.     

It can be shown, that for equidistributed poles such as $\vec{\pole}\iter{0}$ in \eqref{eq:initialGuess}, the far deflation is (see Appendix~\ref{sec:farDeflation})
\begin{equation}
	\deflation{\poleIndex}{ \vec{\pole}\iter{0}_{\textrm{far}}} = \frac{1}{\pole_\poleIndex\iter{0}} \frac{\N - \numNear - 1}{2}.
\end{equation}
Thus, the total deflation may be approximated by
\begin{equation}
	{\approxDeflation{\poleIndex}{\vec{\pole}\iter{\iterIndex}}} = \deflation{\poleIndex}{ \vec{\pole}\iter{\iterIndex}_{\textrm{near}}} + \deflation{\poleIndex}{ \vec{\pole}\iter{0}_{\textrm{far}}}
	\label{eq:approxDeflation}
\end{equation}
if the far poles $\vec{\pole}_{\textrm{far}}\iter{\iterIndex}$ are sufficiently uniformly distributed. By sorting the system poles iterations along pole angles, we can find the near poles $\vec{\pole}_{\textrm{near}}\iter{\iterIndex}$. To establish the quality of this approximation, let us assume that there exists an upper bound $\deflationError$ for the approximation error of the deflation term, i.e.,
\begin{equation}
	\deflationError \geq \max_{\poleIndex,\iterIndex} \abs{\approxDeflation{\poleIndex}{\vec{\pole}\iter{\iterIndex}} - \deflation{\poleIndex}{\vec{\pole}\iter{\iterIndex}}}.
\end{equation}
We can show that the error tolerance $\tol_3$ in \eqref{eq:EAIerror} is satisfied if (see Appendix~\ref{sec:deflationError})
\begin{equation}
\abs{\newton{\pole_\poleIndex\iter{\iterIndex}}^{-1} -  \approxDeflation{\poleIndex}{\vec{\pole}\iter{\iterIndex}}} - \deflationError \geq \frac{2}{\tol_3}. 
	\label{eq:EAIerrorSufficient}
\end{equation}
In other words, if the deflation approximation is sufficiently far from the inverse Newton term, the deflation error becomes negligible. On the contrary, if the deflation term is close to the inverse Newton step, the EAI step error can be large even for small deviations in the deflation term. In case the error tolerance in \eqref{eq:EAIerrorSufficient} is not satisfied, we compute the exact deflation instead.

To implement the proposed approximate deflation, we need a priori knowledge of the approximation error bound $\deflationError$. As $\deflationError$ depends on many factors such as matrix size $\matSize$, system order $\N$, feedback matrix $\Fbm$ and number of near poles $\numNear$, in this work, it is determined experimentally from random FDNs. The performance of the approximate deflation method may be quantified by the number of exact deflations versus approximate deflations. As the initial estimates $\vec{\pole}\iter{0}$ are equidistributed, $\vec{\pole}\iter{1}$ may be computed only from the estimated far deflation with $\numNear = 0$.

\subsection{Residues}
\label{sec:Residues}
Once we have found the system poles, the residues of the modal decomposition \eqref{eq:modalDecomposition} are computed by
\begin{equation}
	\residue_\poleIndex = \frac{\gcq(\pole_\poleIndex)}{\gcp'(\pole_\poleIndex)},
	\label{eq:residueDefinition}
\end{equation}
where we assume that all poles are unique. Similar, but more intricate solutions exists for non-unique poles \cite{Ma:2014jb}. The undriven residue, i.e., the system response without excitation, is
\begin{equation}
	\unresidue_\poleIndex = \frac{1}{\gcp'(\pole_\poleIndex)}.
\end{equation}
The undriven residue is a valuable intermediate step to analyze the mode initial amplitude independent from the input and output drives $\gcq(\pole_\poleIndex)$. Since, $\MatPoly(\pole_\poleIndex)$ is a singular matrix, the derivative of the GCP $\gcp'(\pole_\poleIndex)$ in \eqref{eq:detDerivative} may only be computed by the adjugate formulation. Since $\detp{ \MatPoly(\pole_\poleIndex)} = 0$, the input-output drives in \eqref{eq:gcq} are 
\begin{equation}
	\gcq(\pole_\poleIndex) = \Outgain\tran \, \adj( \MatPoly(\pole_\poleIndex) ) \,  \Ingain.
	\label{eq:residueDrives}
\end{equation}
The difference between the driven and undriven residues may be expressed as a linear combination of the  matrix entries of $\adj(\MatPoly(\pole_\poleIndex))$. Alternatively, the driven residues may also be computed by a least linear squares fit for the time-domain impulse response since the sum of complex resonators in \eqref{eq:sumModes} depends linearly on the residues \cite{Bank:2018il}. 


\subsection{Polynomial Feedback and Delay Matrices}
Although, the focus of this work is on frequency-independent feedback matrices $\Fbm$, much of the development in Section~\ref{sec:numericalModalDecomposition} is applicable to general polynomial matrices. Therefore, it is easy to include further filtering such as a frequency-dependent feedback matrix $\Fbm(z)$. There also exists a singular value decomposition for polynomial matrices $\Fbm(z)$ \cite{Foster:2010ho}. Alternatively, the delay lines are often extended with an attenuation or allpass filter $\alpha_i(z)$, i.e.,
\begin{equation}
	\stdDelay \mat{\alpha}(z) = \diag{ z^{-m_1} \alpha_1(z), \dots, z^{-m_N} \alpha_N(z) }.
	\label{eq:delayLineAbsorption}
\end{equation}
It is important to note that additional filters may increase the  number of system poles. Further, if $\matPoly(z)$ is a rational polynomial, in other words consists of IIR filters, then the transfer function in \eqref{eq:transferFunction} is no longer proper, i.e., the polynomial degree of the nominator is larger than the polynomial degree of the denominator \cite{Shahrrava:2018kw}. Nonetheless, improper partial fraction decomposition can be solved with a delayed parallel form by separating the FIR and IIR part of the transfer function \cite{Bank:2018il}.

For a unitary feedback matrix $\Fbm$ and for attenuation filters in \eqref{eq:delayLineAbsorption}, it is possible to improve the pole magnitude bounds \eqref{eq:magnitudeBounds} to 
\begin{equation}
	\min \paren*{\mat{\alpha} \paren*{e^{\imath \angle \pole_\poleIndex}}^{1/\Delay}} \leq\abs{\pole_\poleIndex} \leq \max \paren*{\mat{\alpha} \paren*{e^{\imath \angle \pole_\poleIndex}}^{1/\Delay}},
	\label{eq:magnitudeBoundsDiagonal}
\end{equation}
where all vector operations are element-wise (see Appendix~\ref{sec:magnitudeBounds}).

\section{Modal Synthesis and Evaluation}
\label{sec:modalEvaluation}
The following evaluation uses real-valued FDN parameters such that the system poles appear in complex conjugate pairs.

\subsection{Modal Synthesis and Accuracy}
A numerically accurate way to verify the modal decomposition is to synthesize each mode $\tfn{\poleIndex}(n)$ in time-domain as expressed in \eqref{eq:modeExp} and compare the sum of all modes with the impulse response $\tf{}(n)$ computed by the time-domain recursion in \eqref{eq:timeDomainFDN}. The concept of modal synthesis and verification is depicted in Fig.~\ref{fig:ModalDecompositionConcept}. 
The error is given by the maximum difference\footnote{The maximum error is chosen as it is an upper bound for the root mean squared error (RMSE) and as such a strict error measure.} between the two impulse responses, i.e., 
\begin{equation}
	\err = \max_n \abs{ \tf{}(n) - \sum_{\poleIndex=1}^\N \tfn{\poleIndex}(n) }.
\end{equation}
In this work, we use double precision floating point arithmetic and the modal decomposition is regarded successful if the maximum error $\err < 10^{-10}$. The EAI is numerically stable as the matrix inversion in \eqref{eq:EAIsimple} is only necessary if the matrix is sufficiently non-singular due to the first stopping criteria. For large delays $\Delay$, the evaluation of $\stdDelay$ may become extremely large or small if $z$ is too far away from the unit circle. At the same time, the poles tend to be close to the unit circle for large delays $\Delay$ due to the bounds given in \eqref{eq:magnitudeBounds}. As a practical intervention, the pole location is clipped to the magnitude bounds if the EAI step causes the pole location to exceed the bounds.

\begin{figure}[!t]
  \includegraphics[width=0.5\textwidth]{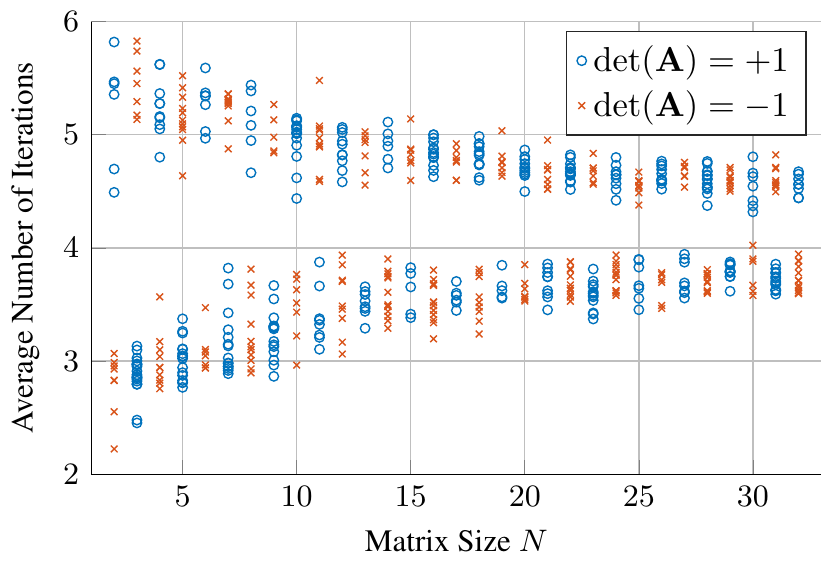}
  \caption{Average number of full iterations in the EAI for 500 random FDNs with total delay $\N$ between 50 and $10^4$ samples and a random orthogonal feedback matrix. The average number of full iterations indicate the average number of Newton steps each pole requires to converge. For low matrix size $\matSize$, the sign of matrix determinant $\detp{\Fbm}$ and parity of $\matSize$ plays a significant role.}
  \label{fig:plot_stepCounter}
\end{figure}

\subsection{Numerical Evaluation}
For the FDN, a single EAI step in \eqref{eq:EAIsimple} can be evaluated in $\BigO(\N + \matSize^3) $: an evaluation of $\MatPoly(z)$ and $\MatPoly'(z)$ is merely an evaluation of the delay matrix $\stdDelay$ in $\BigO(\matSize)$; a numerical matrix inversion can be performed in $\BigO(\matSize^3)$; and the deflation term is evaluated in $\BigO(\N)$. Thus, a full iteration from $\vec{\pole}\iter{\iterIndex}\rightarrow\vec{\pole}\iter{\iterIndex+1}$ can be evaluated in $\BigO(\N^2 + \N\matSize^3)$. This compares favorably with the bound $\BigO(\N^3)$ of a matrix-based algorithm applied to the linearization in \eqref{eq:linearization}. For a high number of system poles $\N \gg N^3$, the complexity of computing the deflation term in \eqref{eq:Deflation} becomes the dominating part. The complexity of the approximate deflation in \eqref{eq:approxDeflation} is similar asymptotically, however in practice, the computational complexity is reduced significantly.

Fig.~\ref{fig:plot_stepCounter} shows the average number of full iterations depending on the matrix size $\matSize$, total delays $\N$ between 50 and $10^4$ samples and a random orthogonal feedback matrix. It can be seen that the number of iterations is largely dependent on the parity of the matrix size and $\detp{\Fbm}$. This illustrates how the initialization \eqref{eq:initialGuess} influences the performance of the EAI. Overall, about 4 to 5 iterations per root may be expected for the EAI to converge. 

Fig.~\ref{fig:plot_eigComparison} depcits a comparison of measured computation time with the MATLAB\footnote{Matlab is a registered trademark of The MathWorks Inc. All computations were performed with Matlab R2016b on a desktop machine with an Intel Core i7 @ 3,40 GHz and 32 GB of RAM.} functions \emph{eig} and \emph{roots} solving the direct problems \eqref{eq:linearization} and \eqref{eq:gcp_pm}, respectively. The total number of delays $\N$ were distributed randomly among eight delay lines and the feedback matrix $\Fbm \in \set{R}^{8 \times 8}$ was a random orthogonal matrix. All methods gave the correct answer with the required accuracy. For the approximate deflation, the number of near poles $\numNear$ was set to $\N / 100$. The maximum deflation error $\deflationError = 10^3$ was determined a priori by probing an independent set of random FDNs of similar configuration. The EAI step tolerance $\tol_3$ was set to $10^{-3}$.

\begin{figure}[!t]
  \includegraphics[]{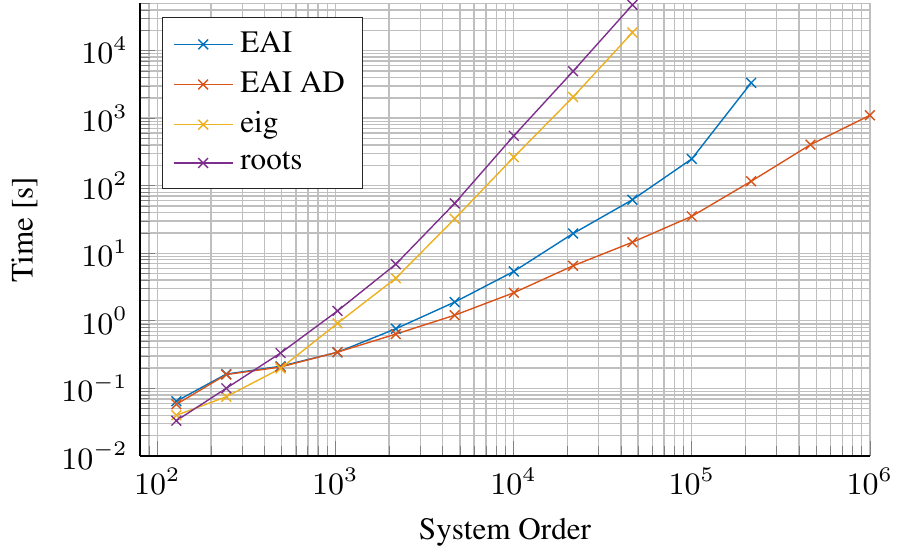}
  \caption{Computation time comparison of EAI with MATLAB build-in functions \emph{eig} and \emph{roots}. For system order $\N > 5 \cdot 10^4$, the memory requirements of \emph{eig} and \emph{roots} become prohibitive on a personal computer configuration. The results are identical to a maximum error less than $\err < 10^{-10}$.}  
  \label{fig:plot_eigComparison}
\end{figure}

The EAI implementation utilizes only standard MATLAB functions and no C-optimization which explains relatively poor performance for small system order $\N < 10^3$. For high system order $\N$ such as $~5 \cdot 10^4$, the standard EAI and EAI with AD outperform the MATLAB's \emph{eig} function by a factor of more than 300 and 1300, respectively. Further, the memory requirements of the EAI are only linear in $\N$ and cubic in $\matSize$ such that it is possible to perform modal decomposition up to $\N=10^6$. Whereas the memory requirements for \emph{eig} become prohibitive for $\N > 5 \cdot 10^4$. For $\N > 10^5$, more the $95\%$ of the computation time of the standard EAI was spent on the deflation term. For the EAI with AD, the number of exact iterations never exceeded 1\% of the total number of iterations proofing the chosen heuristic parameters effective. The EAI with AD performs similar for small delays but outperforms the standard EAI by a factor of 100 for large delays. Each EAI step is independent and only requires synchronization at every full iteration step such that the overall performance of the EAI might further be improved by parallelization.

\section{Analysis of Feedback Delay Networks}
\label{sec:analysisFDN}
We study two applications of modal decomposition in artificial reverberation: Firstly, we study the effect of attenuation filters on the poles and residues of an FDN. Secondly, we study the statistical distribution of poles and residues of random lossless FDNs.

%
%
%

\begin{figure}[!t]
\centering
\subfloat[System pole magnitudes]{%
  \includegraphics[]{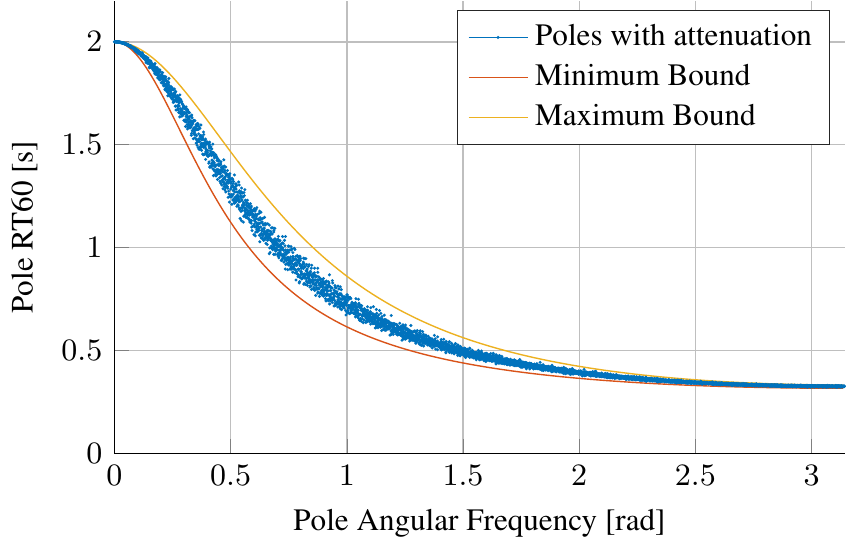}
  \label{fig:plot_onePoleAbsorptionPol}
  }
  
\subfloat[System residue magnitudes]{%
  \includegraphics[]{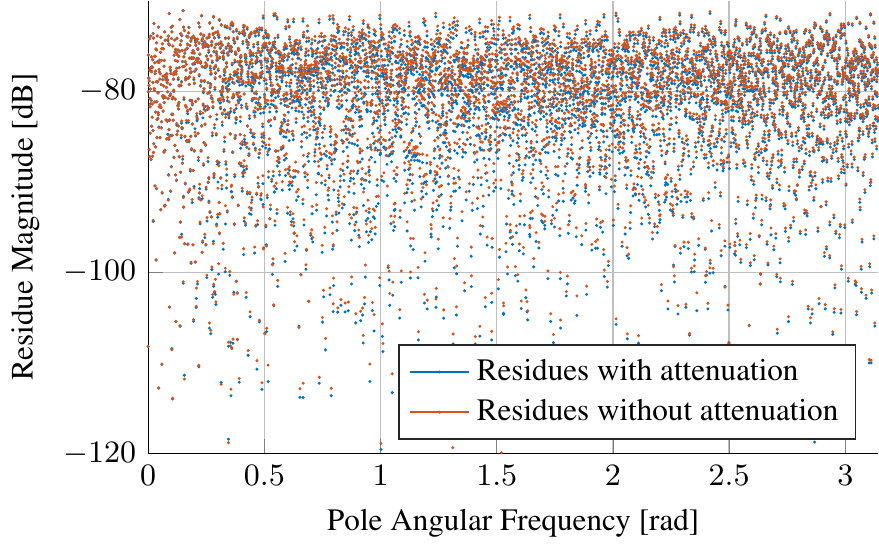}
  \label{fig:plot_onePoleAbsorptionRes}
  } 
  \caption{Modal Decomposition of 8-FDN with target reverberation time $\RT(0) = 2$~seconds and $\RT(\pi) = 0.4$~seconds using one-pole attenuation filters \cite{Jot:1991tq}. Delays are $\Delay = \bracket*{2300,499,1255,866,729,964,1363,1491}$ and $\Fbm$ is a random orthogonal matrix. \protect\subref{fig:plot_onePoleAbsorptionPol} Pole magnitudes converted to reverberation time. Minimum and maximum bounds are computed from \eqref{eq:magnitudeBoundsDiagonal}.  \protect\subref{fig:plot_onePoleAbsorptionRes} Residue magnitudes with and without attenuation. The mean difference between the residue magnitudes is 0.48~dB.}
  \label{fig:OnePole}
\end{figure}

\subsection{Attenutation}
Attenuation filters in FDNs, as they are typically applied in artificial reverberation, aim to control the frequency-dependent reverberation time \cite{Jot:1991tq,Schlecht:2017va}. As expressed in \eqref{eq:delayLineAbsorption}, all delays are extended with absorption filters $\mat{\alpha}(z)$ and the feedback matrix $\Fbm$ is orthogonal or more generally unilossless \cite{Schlecht:2017jt}. We study three types of attenuation: homogeneous, near-homogeneous and inhomogeneous attenuation.

\subsubsection{Homogeneous Attenuation}
The attenuation filters $\mat{\alpha}(z)$ are called homogeneous if there exists an attenuation-per-sample $\atten(z)$ such that
\begin{equation}
	\alpha_i(z) = \atten(z)^{\delay_i}.
\end{equation}
The attenuated delay lines can be expressed as plain delay lines with a mapped argument, i.e., 
\begin{equation}
	\stdDelay \mat{\alpha}(z) = \stdDelayArg{z \atten(z)^{-1}}.
\end{equation}
Consequently, the system poles with attenuation $\pole^{\atten}_\poleIndex$ can be related to the system poles $\pole_\poleIndex$ without attenuation by
\begin{equation}
	\pole_\poleIndex = \pole^{\atten}_\poleIndex \atten(\pole^{\atten}_\poleIndex)^{-1}.
\end{equation}
If we assume that the attenuation filters have a purely real frequency response\footnote{Although such a frequency response is not realizable with a digital filter in general, but useful for the theoretical analysis.}, i.e., $\atten(e^{\imath \omega}) \in \set{R}$ then the mode frequencies are unaltered by the attenuation, i.e., $\angle \pole_\poleIndex = \angle \pole^{\atten}_\poleIndex$. For a unilossless $\Fbm$, all unattenuated system poles $\pole_\poleIndex$ lie on the unit circle such that
\begin{equation}
	\abs{\pole^{\atten}_\poleIndex} = \atten(\pole_\poleIndex)
	\label{eq:onlyMagnitudeAttenuation}
\end{equation}
and the attenuated FDN is stable if $\abs{\atten(e^{\imath \omega})} < 1$. For homogeneous attenuation, the magnitude bounds in \eqref{eq:magnitudeBoundsDiagonal} are tight.

\subsubsection{Near-homogeneous Attenuation}
Typically, the attenuation filters are implemented with relatively low order, such as one-pole filters \cite{Jot:1991tq}, however higher order filters were proposed as well \cite{Jot:2015wv,Schlecht:2017va}. The attenuation filters are designed to match the magnitude response
\begin{equation}
	\abs{\alpha_i(e^{\imath \omega})} \approx \abs{\atten(e^{\imath \omega})^{\delay_i}},
	\label{eq:delayProportional}
\end{equation}
where the attenuation-per-sample is derived from a target reverberation time
\begin{equation}
	20 \log_{10} \abs{ \atten \paren*{e^{\imath \omega}}} = \frac{-60}{\RT(\omega) \FS },
\end{equation}
where $\FS$ is the sampling frequency and $\RT(\omega)$ is the time in seconds for the energy decay curve of the impulse response at frequency $\omega$ to decay by 60~dB \cite{Schroeder:1965du}. For illustration, we compute the one-pole filter according to \cite{Jot:1991tq} given the target reverberation time at DC $\RT(0)$ and Nyquist frequency $\RT(\pi)$. Figure~\ref{fig:OnePole} depicts the resulting modal decomposition for an 8-FDN with an orthogonal feedback matrix and a target reverberation time $\RT(0) = 2$~seconds and $\RT(\pi) = 0.4$~seconds. The system pole magnitudes are modified according to the target reverberation time. However, the attenuation varies especially in the transition band due to errors in the magnitude response caused by the limited filter order. The magnitude of the residues are depicted in Fig.~\ref{fig:plot_onePoleAbsorptionRes}. For near-homogeneous attenuation, it can be observed that the residues with and without attenuation are rather similar. Although the attenuation filters are not completely homogeneous, their phase component is small compared to the phase of the delays $z^{-\Delay}$ such that the overall behavior is well approximated by \eqref{eq:onlyMagnitudeAttenuation}. This suggests that studies on residues of lossless systems may translate well to results for moderately lossy systems.

\begin{figure}[!t]
  \includegraphics[]{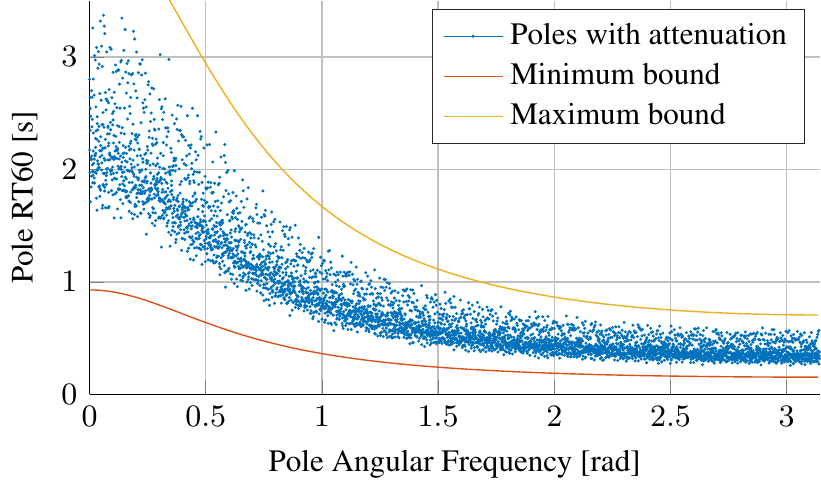}
  \caption{Modal Decomposition of 8-FDN with inhomogeneous attenuation $\mat{\alpha}(z)$ according to an average delay length $\overline{m} = 1074$ for all one-pole attenuation filters in \eqref{eq:delayProportional}. Identical delays, feedback matrix and target reverberation time as in Fig.~\ref{fig:OnePole} were used. Minimum and maximum bounds are computed from \eqref{eq:magnitudeBoundsDiagonal}.}
  \label{fig:inhomogeneousAttenuation}
\end{figure}

\subsubsection{Inhomogeneous Attenuation}
While the homogeneous attenuation has perceptually desirable properties in artificial reverberation, more physically oriented FDN designs such as scattering delay networks \cite{DeSena:2015bb} and radiance transfer \cite{Bai:2015bw} employ attenuation filters which are unrelated to the delay lengths but related to the boundary materials of the simulated space. Figure~\ref{fig:inhomogeneousAttenuation} depicts the modal decay rate of the same 8-FDN as in Fig.~\ref{fig:OnePole} with different attenuation filters. Instead of the delay proportional design in \eqref{eq:delayProportional}, all one-pole filters have the same target frequency response corresponding to an average delay length. As a consequence, the decay time of the neighboring modes are largely different, while the overall shape still follows the target reverberation time. 


\subsection{Statistical Distribution of Poles and Residues}
We present a set of statistical analyses of lossless FDNs which rely on the proposed large-scale numerical computation of the modal decomposition and are difficult to derive by analytic methods. The statistical analysis answers a long-standing question in artificial reverberation design \cite{Karjalainen:2001tp}: \emph{Why do some FDNs have an unpleasant metallic ringing despite a sufficiently high modal density}? While ideal late reverberation has been characterized as Gaussian white noise \cite{Moorer:1979hi}, the metallic ringing is caused by excessive energy at few frequencies. In terms of modal decomposition, metallic ringing may be caused by either clustering of multiple poles at the ringing frequencies or largely varying energy of neighboring modes. We study the following two questions:
\setlist[enumerate]{noitemsep,label*=\arabic*)}
\begin{enumerate}
	\item What is the distribution of the mode frequencies? 
	\item What is the distribution of residue magnitudes? 
\end{enumerate} 
In the analyses, we rely on Monte Carlo simulations of randomly generated lossless FDNs.


\subsubsection{Mode Frequency Distribution}
The near-equidistribution of mode frequencies has been conjectured before \cite{SmithIII:2010vg} and the authors have given an analytical bound on the equidistribution based on Hayman's theorem \cite{Schlecht:2015hi}. The cluster number 
 \begin{equation}
 	\clusterNumber(\omega) = \# \Set*{ i ; \angle \pole_\poleIndex \in \bracket*{\omega - \frac{\pi}{\N}, \omega + \frac{\pi}{\N}} },
 \end{equation}
 is a measure on how equally distributed the mode frequencies are. Here, $\#$ denotes the cardinality of a set. The higher the cluster number, the more poles cluster around the frequency $\omega$. In contrast, a mode gap occurs if $\clusterNumber(\omega) = 0$, i.e., no mode lies in this frequency interval. For perfectly equidistributed poles $\clusterNumber(\omega) = 1$ for all $\omega$. We evaluate the distribution of mode frequencies by computing the histogram of cluster numbers
 \begin{equation}
 	\hist( \histIndex ) = \sum_{l = 1}^{\clusterProbe} \delta \paren*{ \clusterNumber\paren*{ \frac{2 \pi l}{\clusterProbe}  } - \kappa }, 
 \end{equation}
 where $\delta( \cdot )$ is the dirac function, $\histIndex$ is the integer cluster size, and $\clusterProbe$ is the number of observations. For large enough $\clusterProbe$, the histogram converges towards the probability of cluster numbers. The random 8-FDNs have delays between 50 and 1000 samples and an orthogonal feedback matrix. The probabilities are averaged over 100 random instances each.

 In Table~\ref{table:clusterNumbers}, the probability of cluster numbers for randomly generated FDNs are compared to cluster numbers of a pseudo-uniform random number generator with equal sampling size. The discrepancy of the cluster number from an equidistribution is relatively low for the FDN modes compared to the random number generator. In fact for FDNs, it is very rare to find an interval of width $2\pi / \N$ with more than two modes. In stark contrast, acoustic mode density of physical spaces increase quadratically with frequency \cite{Kuttruff:2009vl}.

\begin{table}[!t]
\renewcommand{\arraystretch}{1.3}
\caption{Probability $\hist(\histIndex)$ of cluster numbers of mode frequencies}
\label{table:clusterNumbers}
\begin{tabularx}{0.47\textwidth}{@{}l*{10}{C}c@{}}
\toprule
Cluster size $\histIndex$ & 0 & 1 & 2 & 3 & $\geq$ 4 \\
\midrule
Uniform Random  	& 0.3690  &  0.3661  &  0.1854  &  0.0610  &   0.0186 \\
Lossless 8-FDN 	& 0.1694  &  0.6632  &  0.1653  &  0.0020  & 0.0001 \\
Equidistributed 	& 0.0000  &  1.0000  &  0.0000  &  0.0000  & 0.0000 \\
\bottomrule
\end{tabularx}
\end{table}

\subsubsection{Residue Magnitude Distribution}
In Section~\ref{sec:Residues}, we have presented the computation of the mode residues for a given set of system poles.  Figure~\ref{fig:residueDistribution} depicts the magnitude histogram of the total and undriven residues as well as the input-output drives for a random 8-FDN. The input-output drives are comprised of all individual input-output combinations, i.e., $\adj(\MatPoly(\pole_\poleIndex))$ in \eqref{eq:residueDrives}. The total residues $\residue(\pole_\poleIndex)$ result from unit input and output gains, i.e., $\Ingain = \ones$ and $\Outgain = \ones$, or in other words, $\residue(\pole_\poleIndex) = \unresidue(\pole_\poleIndex) (\ones\tran \adj(\MatPoly(\pole_\poleIndex)) \ones)$. The magnitude distributions of the inverse undriven residues $1 / \unresidue(\pole_\poleIndex)$, the total residues $\residue(\pole_\poleIndex)$ and input-output drives $\adj(\MatPoly(\pole_\poleIndex))$ all resemble log-Rayleigh distributions \cite{Rivet:2007kx}. However, just by altering the feedback matrix $\Fbm$, it is possible to encounter various other distributions of the residue magnitude.  Figure~\ref{fig:plot_randomLosslessResidues}, depicts the residue magnitude distribution of four selected orthogonal feedback matrices.

\begin{figure}[!t]
  \includegraphics[]{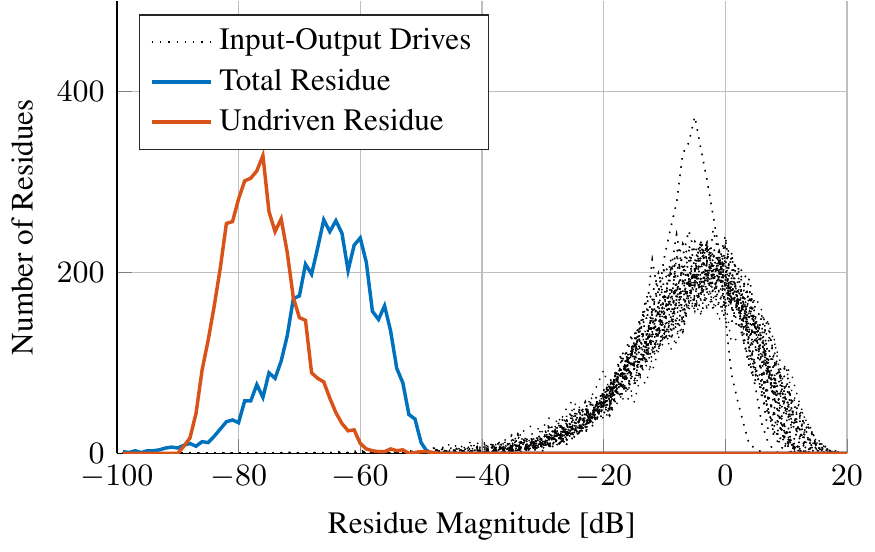}
  \caption{Histogram of residue magnitude of an 8-FDN with delays $\Delay = [492,794,1849,1855,1155,1090,78,1957]$ and a random orthogonal feedback matrix $\Fbm$. The undriven residues $\unresidue(\pole_\poleIndex)$ are dependent only on the feedback loop $\MatPoly(z)$, whereas the total residues $\residue(\pole_\poleIndex)$ results from unit input and output gains, i.e., $\Ingain = \ones$ and $\Outgain = \ones$, respectively. The input-output drives $\gcq(\pole_\poleIndex)$ are the $8 \times 8$ magnitudes of the adjugate matrix $\adj(\MatPoly(\pole_\poleIndex))$.}
  \label{fig:residueDistribution}
\end{figure}

\subsubsection{Discussion}
For randomly generated FDNs, the mode frequencies are nearly equidistributed such that every frequency band has energy contributions from a similar number of modes. On the other hand, the high dynamic range of the residue magnitudes suggest that a small number of poles contribute a large portion of the impulse response energy. In the context of artificial reverberation, the high-energy modes dominate the frequency spectrum such that the audible modal density is considerably lower than theoretic modal density, i.e., the number of modes per frequency. For illustration, we have synthesized audio examples from the four instances depcited in Fig.~\ref{fig:plot_randomLosslessResidues} and provided them  online\footnote{\url{www.audiolabs-erlangen.de/resources/2018-IEEE-Modal}}.   

The residue distribution may be optimized in two steps: Firstly, optimization of the undriven residues by choosing delays $\Delay$ and feedback matrix $\Fbm$. Secondly, optimization of the total residue by choosing the input and output gains, $\Ingain$ and $\Outgain$. While the first step is a non-linear process which requires further research, the second step may be readily solved by linear least square fitting.

\section{Conclusion}
We presented a numerically efficient technique for modal decomposition of the FDN. Standard methods such as eigenvalue decomposition of the linearized system and polynomial root finding methods applied to the characteristic polynomial require significant computational resources when the system order is large. The proposed method applies the Ehrlich-Aberth Iteration to the polynomial matrix formulation of the FDN. Further we proposed, an efficient approximate deflation technique based on the estimation of far poles. For high system order such as $~5 \cdot 10^4$, the standard EAI and approximate EAI outperform the MATLAB's \emph{eig} function by a factor of more than 300 and 1300, respectively. The approximate EAI was able to give reliable results up to a system order of 1 million. The modal decomposition was applied to FDNs in the context of artificial reverberation. Three types of attenuation were studied: homogeneous, near-homogeneous and inhomogeneous. The potential for explicit analysis of the pole and residues was demonstrated for attenuation filter design. Statistical analysis showed that for randomly generated FDNs, the mode frequencies are nearly equidistributed and the residue magnitudes follow a log-Rayleigh distribution. This analysis suggests that relatively few modes are contributing a large portion of the late reverberation energy.

\begin{figure}[!t]
  \includegraphics[width=0.5\textwidth]{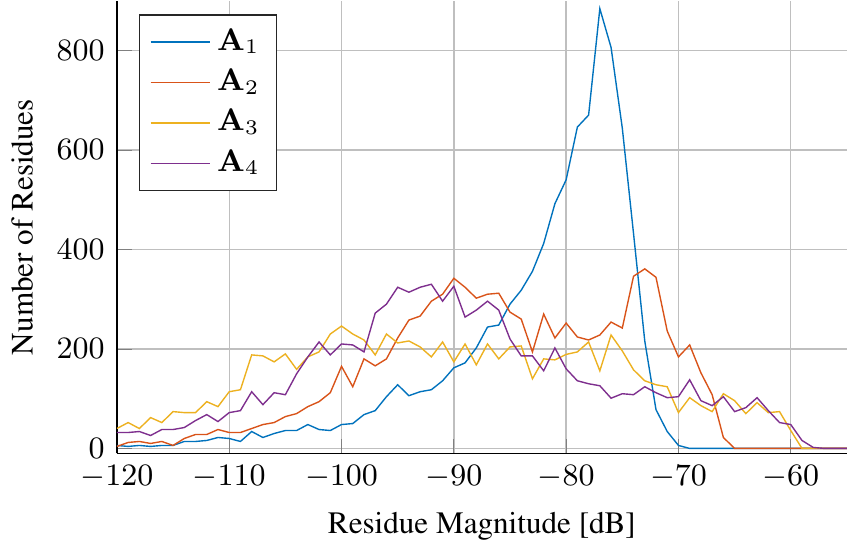}
  \caption{Histograms of the residue magnitude with delays $\Delay = [492,794,1849,1855,1155,1090,78,1957]$ for four different orthogonal matrices $\Fbm_1$, $\Fbm_2$, $\Fbm_3$ and $\Fbm_4$. The four matrices are chosen manually from 1000 random orthogonal matrices to display a variety of residues magnitude distributions.}
  \label{fig:plot_randomLosslessResidues}
\end{figure}

\appendices

\section{Lower bound of Pole Magnitude}
\label{sec:magnitudeBounds}
We present lower and upper bounds on the pole magnitudes $\abs{ \vec{\pole} }$ of an FDN. The bounds are based on the generalization of Rouch\'e's theorem to matrix polynomials. 

\begin{theorem}[see \cite{Monden:1980bo}] Let $\mat{S}(z)$ and $\mat{Q}(z)$ be matrix polynomials and let $r$ be a positive real number . If $\mat{S}(z)\herm \mat{S}(z) - \mat{Q}(z)\herm \mat{Q}(z)$ is positive definite for $|z| = r$, then the polynomials $\detp{ \mat{S}(z)}$ and $\detp{\mat{S}(z) + \mat{Q}(z)}$ have the same number of roots of modulus less than $r$.
\end{theorem}

An immediate consequence of the above theorem applied to the polynomial $\MatPoly(z)$ of \eqref{eq:loopTransfer} with $\mat{S}(z) = -\Fbm$ and $\mat{Q}(z) = \stdDelay^{-1}$ is \cite{Bini:2013fo}: If
\begin{equation}
	\Fbm\herm \Fbm - \stdDelayArg{{z}^\conj}^{-1} \stdDelay^{-1} \succ 0, \textrm{ for } \abs{z} = r
	\label{eq:Rouche}
\end{equation}
where $\mat{X} \succ \mat{Y}$ means that $\mat{X} - \mat{Y}$ is positive definite, then $\MatPoly(z)$ has no eigenvalues in the open disk with center $0$ and radius $r$. The criterium in \eqref{eq:Rouche} is equivalent to
 \begin{equation}
 	\Fbm\herm \Fbm \succ \stdDelayArg{r^{-2}}
 \end{equation}
which in turn is equivalent to \cite{Horn:2013tf}
\begin{equation}
	\spectralRadius \paren*{ (\Fbm\herm \Fbm)^{-1} \stdDelayArg{r^{-2}} } \leq 1, 
	\label{eq:spectralRadiusCrit}
\end{equation}
where $\spectralRadius(\mat{X})$ denotes the spectral radius of a matrix $\mat{X}$. Using properties of the spectral norm \cite{Horn:2013tf} we can give an upper bound on this expression by
\begin{equation}
\begin{aligned}
	\spectralRadius \paren*{ (\Fbm\herm \Fbm)^{-1} \stdDelayArg{r^{-2}} } \leq \euclidean*{(\Fbm\herm \Fbm)^{-1} \stdDelayArg{r^{-2}}} \\
	\leq \euclidean*{\Fbm^{-1}} \euclidean*{\Fbm^{-H}} \euclidean*{ \stdDelayArg{r^{-2}}} \\
	= \euclidean*{\Fbm^{-1}}^2 \euclidean*{ \stdDelayArg{r^{-2}}}.
\end{aligned}
\end{equation}
Thus, the criterium \eqref{eq:spectralRadiusCrit} is satisfied if
\begin{equation}
	 r^{2 \min \Delay} = \euclidean*{ \stdDelayArg{r^{-2}}}  \leq  \euclidean*{\Fbm^{-1}}^{-2} = \min \sv(\Fbm)^2.
\end{equation}
Therefore with \eqref{eq:Rouche}, the pole magnitude lower bound may be given as 
\begin{equation}
	\min \abs{\vec{\pole}} \geq \min \sv(\Fbm)^{1 / \min \Delay}.
\end{equation}
Analogously, applying the same arguments to the reversed matrix polynomial $z^{\max \Delay}\MatPoly(z^{-1})$ yields an upper bound
\begin{equation}
	\max \abs{\vec{\pole}} \leq \max \sv(\Fbm)^{1 / \max \Delay}.
	\label{eq:upperBound}
\end{equation}
For additional attenuation filters $\mat{\alpha}(z)$ as in \eqref{eq:delayLineAbsorption} and a unitary feedback matrix $\Fbm$, these bounds can be tightened further. Rouch\'e's criterion \eqref{eq:Rouche} gives the relation
\begin{equation}
	\mat{\alpha}(z) \succ \abs{z}^{\Delay}.
\end{equation}
Thus, the lower bound of the pole magnitude is
\begin{equation}
	\abs{\pole_\poleIndex} \geq \min \paren*{\mat{\alpha} \paren*{e^{\imath \angle \pole_\poleIndex}}^{1/\Delay}}.
\end{equation}
The corresponding upper bound may be derived similar to \eqref{eq:upperBound}:
\begin{equation}
	\abs{\pole_\poleIndex} \leq \max \paren*{\mat{\alpha}\paren*{e^{\imath \angle \pole_\poleIndex}}^{1/\Delay}}.
\end{equation}
These bounds are tight for a diagonal matrix $\Fbm$.

\section{Far Deflation Estimation}
\label{sec:farDeflation}
We are given the equidistributed poles $\vec{\pole}\iter{0}$ as defined in \eqref{eq:initialGuess} and an even number of near poles $\numNear$. We compute the far deflation for pole $\pole_\poleIndex\iter{\iterIndex}$. First we state a useful identity. For any real $x$, 
\begin{equation}
	\frac{1}{1 - e^{\imath x}} + \frac{1}{1 - e^{-\imath x}} = 1.
\end{equation}
The total deflation is 
\begin{equation*}
\begin{aligned}
	\deflation{\poleIndex}{ \vec{\pole}\iter{0}} &= \sum_{l=1, l\neq j}^\N \frac{1}{\pole_\poleIndex\iter{0} - \pole_l\iter{0}} \\
	&= \frac{1}{\pole_\poleIndex\iter{0}} \sum_{l=1, l\neq j}^\N \frac{1}{ 1 - \pole_l\iter{0}/\pole_\poleIndex\iter{0} } \\
	&= \frac{1}{\pole_\poleIndex\iter{0}} \sum_{l=1, l\neq j}^\N \frac{1}{ 1 - \exp(\imath2\pi \frac{l-j}{\N}) } 
	= \frac{1}{\pole_\poleIndex\iter{0}} \frac{\N-1}{2}.
\end{aligned}
\end{equation*}
Similarly, as each conjugate pair of poles contribute equally to the deflation, the far deflation is
\begin{equation}
	\deflation{\poleIndex}{ \vec{\pole}\iter{0}_{\textrm{far}}} = \frac{1}{\pole_\poleIndex\iter{0}} \frac{\N - \numNear - 1}{2}.
\end{equation}

\section{Deflation Error}
\label{sec:deflationError}
We show that inequality \eqref{eq:EAIerrorSufficient} is sufficient for inequality \eqref{eq:EAIerror}. For the sake of brevity, we omit the pole arguments in the following. Given the deflation approximation $\approxDeflationS{\poleIndex}$, which satisfy \eqref{eq:EAIerrorSufficient}, we obtain
\begin{equation}
	\frac{\tol_3}{2} \geq \frac{1}{\abs{\newtonS^{-1} -  \approxDeflationS{\poleIndex}} - \deflationError} \geq 0.
	\label{eq:errorDefApprox}
\end{equation}
We show that \eqref{eq:errorDefApprox} satisfies EAI step error tolerance \eqref{eq:EAIerror}. Because $\deflationError \geq 0$, it is
\begin{equation}
	\frac{\tol_3}{2} \geq \frac{1}{\abs{\newtonS^{-1} -  \approxDeflationS{\poleIndex}}} \geq 0.
\end{equation}
Further, as $\abs{\deflationS{\poleIndex} - \approxDeflationS{\poleIndex}} - \deflationError \leq 0$, 
\begin{equation*}
\begin{aligned}
	\frac{1}{\abs{\newtonS^{-1} - \deflationS{\poleIndex}}} 
	&\leq \frac{1}{\abs{\newtonS^{-1} - \deflationS{\poleIndex}} + \abs{\deflationS{\poleIndex} - \approxDeflationS{\poleIndex}} - \deflationError} \\
	&\leq \frac{1}{\abs{\newtonS^{-1} - \approxDeflationS{\poleIndex}} - \deflationError} \leq \frac{\tol_3}{2}	
\end{aligned}
\end{equation*}
Eventually, we can show that
\begin{equation*}
\begin{aligned}
	\abs{ \EAIstep{\poleIndex}{\iterIndex}  -
	\approxEAIstep{\poleIndex}{\iterIndex} } 
	&\leq \abs{ \frac{1}{{\newtonS^{-1} - \deflationS{\poleIndex}}} - \frac{1}{{\newtonS^{-1} - \approxDeflationS{\poleIndex}}}} \\
	&\leq \abs{ \frac{1}{{\newtonS^{-1} - \deflationS{\poleIndex}}}} + \abs{\frac{1}{{\newtonS^{-1} - \approxDeflationS{\poleIndex}}}}
	\leq \tol_3	
\end{aligned}
\end{equation*}

\bibliographystyle{IEEEtran}
\bibliography{Papers}

\end{sloppy}
\end{document}